\documentclass[aps,pra,twocolumn,showpacs,superscriptaddress,groupedaddress]{revtex4}
\usepackage{graphicx}
\usepackage{amssymb}
\usepackage{amsmath}
\usepackage{bm}
\usepackage{color}
\usepackage{paralist}

\begin{document}

\title{Azimuthons and pattern formation in annularly confined exciton-polariton Bose-Einstein condensates}

\author{Guangyao Li}
\affiliation{Nonlinear Physics Centre, Research School of Physics and Engineering, The Australian National University, Canberra ACT 0200, Australia}

\begin{abstract}
We present numerical analysis of steady states in a two-component (spinor) driven-dissipative quantum fluid formed by condensed exciton-polaritons in an annular optically induced trap. We demonstrate that an incoherent ring-shaped optical pump creating the exciton-polariton confinement supports the existence of stationary and rotating azimuthon steady states with azimuthally modulated density. Such states can be imprinted by coherent light pulses with a defined orbital angular momentum, as well as generated spontaneously in the presence of thermal noise.
\end{abstract}

\maketitle
\section{Introduction}

Dynamics of open-dissipative exciton-polariton condensates in optically defined trapping potentials has developed into an active area of research due to high degree of flexibility and scalability afforded by the optical trapping techniques \cite{CarusottoREV13,SchneiderArxiv15}. Annular confinement, in particular, is capable of supporting superfluid polariton currents, which are of potential use for proposed interferometry and sensing devices based on microcavity polaritons \cite{ByrnesREV14}. The annular polariton flow, its stability and disruption \cite{ManniPRL11,DreismannPNAS14,LiPersistentCurrent}, as well as its connection to nontrivial vortices in two-component (spinor) polariton systems \cite{SnokePNAS15} have been vigorously investigated both experimentally and theoretically. The spin degree of freedom of a microcavity polariton is directly linked to the polarization of its photonic component and therefore can be easily mapped out via polarization and frequency-resolved optical tomography of the cavity photoluminescence. This straightforward detection method has enabled a multitude of experimental studies of half-solotons, half-vortices, spin vortices, and other non-trivial spin textures spontaneously occuring in exciton-polariton condensates \cite{SnokePNAS15,LagoudakisScience09,KammannPRL12,HivetNatPhys12,ManniNatComm13,Swirls15}.

The majority of non-trivial spin dynamics in polariton condensates is associated with the Rashba-like coupling between the spin components introduced by the effective magnetic field induced by the momentum-dependent TE-TM energy splitting between the polariton modes \cite{Shelykh05,Kavokin05}. However, spontaneous formation of spin patterns and non-trivial spin dynamics \cite{BerloffPRB10,LiSpin15} can also be caused by the asymmetry-induced, momentum-independent linear coupling between the circular polarization components, which commonly arises due to the strains in the semiconductor heterostructures. 

In this work, we examine non-trivial spin states of the dynamical system describing non-equilibrium, incoherently pumped Bose-Einstein condensate (BEC) of exciton-polaritons trapped by an annular potential induced by the pump. We show, that this pumping configuration supports steady vortex states with azimuthally modulated density (azimuthons) which can be interpreted as stationary spin waves. Steady rotation with THz-range frequency associated with this states results in optical "ferris wheels" \cite{FrankeOpticsExpress07} in the cavity photoluminescence. We also describe the stationary pattern formation supported by nonlinear instabilities of the annular polariton flow, and show that the noise naturally present in the system due to, e.g., thermal effects, allows for spontaneous formation of vortex azimuthons.

\section{The Model}\label{Sec:model}


The mean-field dynamics of a two-component (spinor) polariton condensate can be described by the open-dissipative Gross-Pitaevskii (GP) equation coupled to the rate equations for spin-polarized reservoir of hot 'excitonic' polaritons created and replenished by a non-resonant optical pump \cite{WoutersPRL07,KammannPRL12}. In the circularly polarized basis $\psi_{\pm}$, where $+$ and $-$ stand for the right- and the left-hand circular polarization component respectively, the dynamical model is written as follows ($\sigma=\pm$):
\begin{eqnarray}\label{eq:rescaled}
\begin{aligned}
i\partial_{t}\psi_{\sigma} &=\left\{ -\frac{1}{2}\nabla^{2}+u_{a} \lvert \psi_{\sigma} \rvert^{2}+u_{b} \lvert \psi_{-\sigma} \rvert^{2}+g_R\, n_\sigma \right.\\
                           &\qquad \left.+\frac{i}{2}\left[R\, n_{\sigma}-\gamma_c\right]\right\} \psi_{\sigma}+J\, \psi_{-\sigma}, \\ 
\partial_{t}n_{\sigma}&=P_\sigma(r)-(\gamma_R+R\, \lvert \psi_\sigma \rvert^2)n_\sigma,\\  
\end{aligned}
\end{eqnarray}
where $u_a$ and $u_b$ represent ($|u_b|<|u_a|$ \cite{Vladimirova10}) the same-spin and cross-spin s-wave scattering strengths respectively, $g_R$ characterises interactions between the condensate and reservoir (the blueshift energy), $R$ is the same-spin stimulated scattering rate from the reservoir into the condensate, $\gamma_{c}$ is the loss rate of polaritons with $\gamma_{c}=1/\tau_c$ where $\tau_c$ is the polariton lifetime, and $J$ is the internal Josephson coupling. For the reservoir equation, $n_\sigma$ is the spin-dependent reservoir density \cite{KammannPRL12}, $P_\sigma$ is the spin-dependent pumping rate, and $\gamma_R$ is the loss rate of the reservoir polaritons. The anisotropic TE-TM splitting effect is assumed to be weak and thus not taken into account \cite{GlazovPRB15}. We also assume that the cross-spin stimulated scattering is negligible comparing with the same-spin counterpart \cite{BerloffPRB10}. As shown in \cite{LiSpin15}, weak cross-spin stimulated scattering does not significantly affect the dynamics. 


Equation~(\ref{eq:rescaled}) is written in the dimensionless form by using the characteristic scales of time $\tau_c$, length $L=\sqrt{\hbar/(m\gamma_{c})}$, and energy $E_{u}=\hbar \gamma_c$. We assume parabolic dispersion approximation near the polariton ground state, where $m$ is the effective mass of the lower polaritons. All unspecified parameters in Eq.~(\ref{eq:rescaled}) take the default numerical values in \cite{parameters}. For these parameters, the unit of time, $t=1$, used in dynamical simulations throughout this work, corresponds to $3$ ps.

Although generally the energy functional corresponding to Eq.~(\ref{eq:rescaled}) takes complex values, when the pumping and decay reach equilibrium there exist dynamically stable steady states whose energy functionals are strictly real \cite{LiPersistentCurrent}. It suggests that the condensate dynamics can be approximately characterized by the real part of the energy functional \cite{Bao13}, which is given by $E=E_{+} + E_{-}$, where
\begin{equation}\label{energy}
\begin{aligned}
E_{\sigma}= & \int d^2 {\bm r}\left[ \frac{1}{2} \lvert \nabla \psi_\sigma \rvert^2+\frac{u_a}{2} \lvert \psi_\sigma \rvert^4+g_R\, n_\sigma \lvert \psi_\sigma \rvert^2 \right. \\
										 & \left. +\frac{u_b}{2} \lvert \psi_\sigma \rvert^2 \lvert \psi_{-\sigma} \rvert^2+J\, \text{Re}(\psi_\sigma\psi^*_{-\sigma})\right], 
\end{aligned}
\end{equation}
and the integration is performed over the area where the condensate density is non-negligible.

Under the incoherent pumping conditions, the phase of the pump beam will be lost during the polariton energy relaxation process due to scattering with phonons. The spatial distribution of the condensate, therefore, is controlled by the pump rate $P_\sigma(r)$ which is proportional to the spatial intensity distribution of the pump beam. In this work, we use a Laguerre-Gaussian (LG) beam to form an annular pumping configuration. The pumping power of the beam is normalized by the threshold power for polariton condensation $\bar{P}=P^{max}/ P_{th}$, where $P_{th}=\gamma_R \gamma_c / R$ is the pumping threshold given by the homogeneous pump approximation \cite{WoutersPRL07}, and $P^{max}$ is the peak intensity of the LG beam.
For a spinor system, the intensity of the LG beam is split into each component as $\bar{P}=\bar{P}_{+}+\bar{P}_{-}$. We denote the polarization bias of the pump as $PL=\bar{P}_{-}/\bar{P}$, \emph{e.g.}, for a linearly polarized pump $PL=0.5$, while for a right-handed circularly polarized pump $PL=0$.

\section{Vortex States}\label{Sec:VS}

The full picture of dynamical phenomena described by Eq.~(\ref{eq:rescaled}) is given by the interplay between the nonlinear interactions and the linear coupling. We start the discussion by reviewing some of the existing results as limiting cases of the this dynamical model and then extend our understanding to the more intricate situations.

If $J=0$, and the cross-spin nonlinear interaction is vanishingly small, two polarization components become effectively decoupled from each other and Eq.~(\ref{eq:rescaled}) reduces to two sets of single-component equations. 
To obtain a steady state, one can require the equilibrium between pumping and decay $R\, n_{\sigma}-\gamma_c=0$ and steady reservoir density $\partial_{t}n_{\sigma}=0$. Together these conditions lead to $\lvert \psi_\sigma \rvert^2 \propto P_\sigma$, \emph{i.e.}, the condensate density distribution follows the intensity distribution of the pump, and is therefore azimuthally homogeneous. Under a LG pump $P_\sigma$, the condensate density distribution has an annular shape which can support vortex states. Detailed discussions of the existence and stability properties of single-component vortex states can be found in Ref.~[\onlinecite{LiPersistentCurrent}]. These states are modulationally stable in certain parameter regimes, and in what follows we will consider only these regimes.

The above conclusion remains valid even in the presence of the cross-spin interaction \cite{BerloffPRB10}. 
Thus, the topological charge of vortex states for each polarization component can be different from each other. 
When the condensate is pumped by a linearly polarized ($PL=0.5$) LG beam and one component acquires non-zero angular momentum, \emph{e.g.}, $m_{+}=1$ and $m_{-}=0$, where $m$ is the topological charge \cite{LiPersistentCurrent}, the system forms the so called half-vortex state \cite{RuboPRL07,RuboPRB14}, which has been observed in experiment \cite{LagoudakisScience09}. When viewed in the linearly polarized basis, such a state forms a rotating vortex with $\pi$ phase jump around the azimuthal coordinate and a density dip \cite{ManniNatComm12}, where the horizontal and vertical polarization components in the linearly polarized basis are defined as
\begin{equation}\label{Linear_Basis}
\psi_H  = \frac{1}{\sqrt{2}} (\psi_{+} + \psi_{-})\quad  \text{and}\quad    \psi_V  = \frac{i}{\sqrt{2}} (\psi_{-} - \psi_{+}).
\end{equation}

\begin{figure}[t]
\includegraphics[width=8cm]{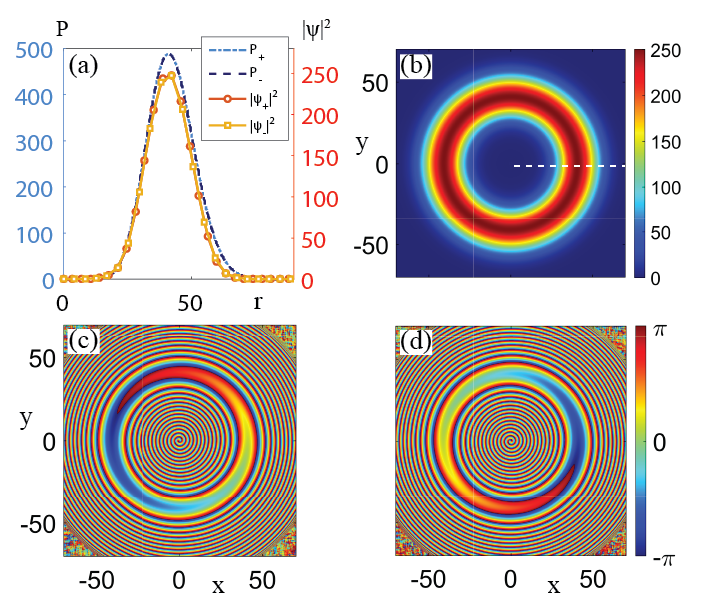} 
\caption{(Color online). Anti-bonding states formed under an LG$_{50}$ mode pumping. (a) Radial profiles of the condensate density and pumping rate along the dashed line in (b). (b) density and (d) phase distribution of the $+$ component. (c) Phase distribution of the $-$ component. Parameters: $\bar{P}=6, PL=0.5$, and $J=0.5$.}
\label{LG_linear_Pump}
\end{figure}

If $J \neq 0$ (without lost of generality we assume $J>0$ throughout this work), the linear coupling induces polariton density exchange between two components, the so-called Josephson currents (see Eq.~(\ref{eq:I_J})), which introduces spin dynamics into our system \cite{WoutersPRB08}. Assuming that the spatial variation of the condensate can be ignored (the homogeneous approximation), previous studies have shown that, under a linearly polarized ($PL=0.5$) pump, with sufficiently large $J$ the condensate will fall into the anti-bonding state, where the relative phase between the two components maintains $\pi$ difference \cite{LiSpin15}. 
Now we model the pumping configuration with a linearly polarized LG beam whose cross-section profile is shown in Fig.~\ref{LG_linear_Pump}(a). 
The condensate falls into the anti-bonding state with spatially inhomogeneous density distribution. The relative phase between the two components maintains $\pi$ difference over the whole pumping region, and if one component
forms a vortex state with the topological charge $m=1$, 
the fixed phase difference will drag the other component into the same vortex state with an overall $\pi$ phase lapse, see Fig.~\ref{LG_linear_Pump}. This is a static half-vortex state: when viewed in the linearly polarized basis the state does not rotate. 

The above two cases are two-component stable vortex states given by simple combination of previous results, and they form the qualitative description of our system. When $J$ is weak, two components are loosely coupled; and when $J$ is large, they maintain a fixed relative phase, regardless of their spatial distribution. The angular momentum acquired by each component can be imprinted by a coherent LG pulse in the initial state \cite{SanvittoNatPhys10}. The coherent phase imprinting enforces formation of pre-determined vortex states. However, as demonstrated in Sec.\ref{Sec:White_noise}, with a suitable pumping configuration, vortex states can also form spontaneously from white noise.

\section{Vortex azimuthon states}\label{Sec:STstates}

Vortex states in the annular trap created by the optical LG pump have azimuthally symmetric density distributions and azimuthally linear phase distributions over the pumped area. It has been shown previously, both for the conservative GP model, as well as the Nonlinear Schr\"{o}dinger and Ginsburg-Landau models in optics, that vortex states are special cases of more general steady states with periodic azimuthal density modulations that have been realized in optics as azimuthons and in atomic BEC system as soliton train (ST) states.  
Various types of azimuthons have been studied extensively \cite{DesyatnikovPRL05,LopezOpticsExpess05,LashkinPRA09} and have been observed in experiments \cite{MinovichOptExpress09,IzdebskayaOptExpress11}. The notion of vortex azimuthons has recently been extended to open-dissipative systems \cite{LobanovOptExp11,BorovkovaOptExp11,BorovkovaPRA12,OstrovskayaPRA12}.  For conservative (atomic) BEC systems, the analytical expression for the ST state was first developed in \cite{CarrPRA00,CarrPRA00II}. Since then the ST states have been considered both for the single-component case \cite{KanamotoPRA09} and for the two-component case \cite{SmyrnakisPRA13,KamchatnovPRA14}.
Note that in literature the term "soliton train" might refer to a series of propagating solitons under the open-boundary condition \cite{PinskerPRL14}. Here the ST state refers to the one \cite{KanamotoPRA09,SmyrnakisPRA13} with the periodic boundary condition. In the following, within the scope of our discussions,
we will not discriminate between the azimuthon and the ST state. (Detailed comparisons can be found in \cite{DesyatnikovPRL05, KanamotoPRA09}.)

Stable ST states in a single-component polariton system under the incoherent LG pumping scheme is not possible as a result of driven-dissipative nature of the system. As mentioned above, a steady-state condensate and reservoir density distribution should be proportional to the pump rate, \emph{i.e.}, for an annular azimuthally homogeneous pump their density should be azimuthally homogeneous. This argument no longer holds true if the system supports Josephson vortices \cite{KaurovPRA05,SuPRL13,SuPRA15,RoditchevNatPhys15} that stem from internal Josephson currents between the two condensate components. In the simplest case, a Josephson vortex will introduce a sine-shape spatial distribution of Josephson current between two components \cite{SuPRL13}. If the Josephson vortex does not fully compensate the density difference between the two components, one can expect that the density modulation of a vortex state will form cnoidal waves \cite{AtrePRE06} that mimic the conservative soliton state.

The particle density imbalance can be introduced by a spin-biased pump. The homogeneous spin dynamics considered in \cite{LiSpin15} dictates that, if the polarization of the pump slightly deviates from the linear one ($PL=0.5$), the condensate will still form a steady state with a fixed relative phase which is close to but not exactly $\pi$. We denote such a state as semi-anti-bonding (SAB) state and the corresponding relative phase is denoted as $\theta_{s}$. The relative phase between $\psi_{+}$ and $\psi_{-}$ is defined by
\begin{equation}\label{eq:theta} 
\theta (\textbf{r},t)=\phi_{-} (\textbf{r},t) -\phi_{+} (\textbf{r},t),
\end{equation}
where $\phi_\pm$ is the phase of $\psi_\pm$.
In such a state, the Josephson current maintains the relative phase $\theta_{s}$ between two components throughout the whole pumped area. In an annular pumping configuration, if both components acquire non-zero angular momentum $m_\pm$ (not necessarily the same), the spatial flow of the condensate will lead to spatial variation of the phase in each polarization component. This variation is governed by the relation $\textbf{v} \sim \nabla \phi$, where $\textbf{v}$ is the velocity of the condensate flow and $\phi$ is the phase, and will lead to the deviation of the  relative phase between the components from $\theta_{s}$. The competition between the azimuthal flow governed by the nonlinear interactions within each component and the Josephson current given by the linear coupling between the two components results in cnoidal rotating waves that are very similar to that of the ST states found in atomic BEC systems.

\begin{figure}
\includegraphics[width=9cm]{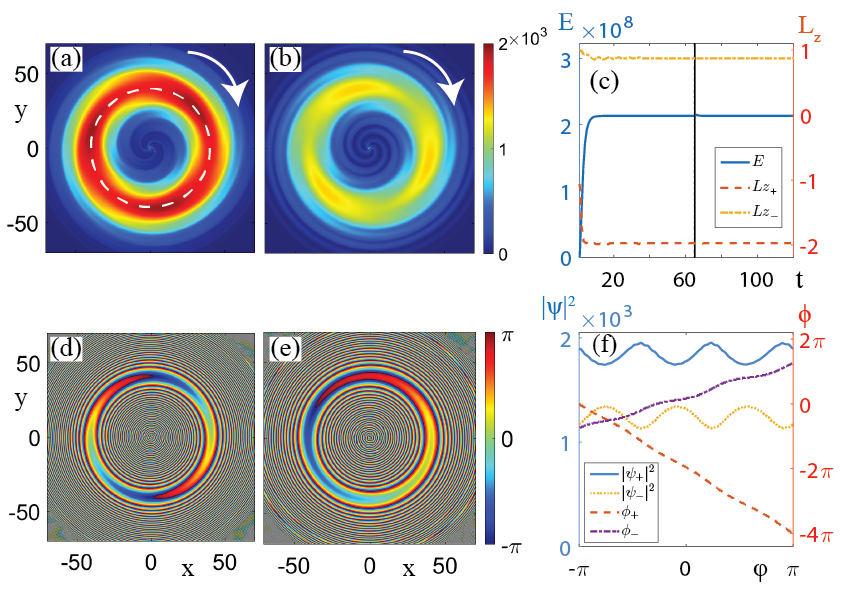} 
\caption{(Color online). Spatial distribution and time evolution of a soliton train state. 
(a)-(b) density and (d)-(e) phase distribution for $\psi_+$ (a) (d) and $\psi_{-}$ (b) (e).
White arrows indicate the rotation direction of the density dips. (c) Time evolution of the energy and normalized orbital angular momentum. Vertical dotted line: perturbation added time. (f) Azimuthal distribution of density and phase along the dashed line in (a). Parameters: $\bar{P}=25, PL=0.4, m_{+}=-2, m_{-}=1$,  and $J=0.17$.}
\label{STstates}
\end{figure}

Fig.~\ref{STstates} shows a snapshot of the ST state. In this case, the $+$ component was pumped stronger than the $-$ component, which is demonstrated by the pseudo-color representation of the polariton density. Although each component acquired different angular momentum, $m_{+}=-2$ and $m_{-}=1$, their densities rotated in the same direction as indicated by the white arrows. 
Density dips can be seen in their density distribution for both components. 
The number of density dips is given by the phase winding difference between the two components, and in the current case specifically $j=|m_{+}-m_{-}|=3$. The ST state is spatially inhomogeneous and the dimensionality reduction method used in \cite{LiPersistentCurrent} is no longer applicable. Nevertheless, the condensate can be qualitatively represented by the area pumped most strongly by the LG beam, as indicated by the white dashed-ring in Fig.~\ref{STstates}(a). Fig.~\ref{STstates}(f) shows the density and phase distribution around the ring for both components. The Josephson current cannot fully compensate the density difference around the ring, and the azimuthal density dips distribution in the two components are complementar and are associated with the azimuthally nonlinear distribution of phase. Fig.~\ref{STstates}(c) further demonstrates that the ST state is stable to a weak broadband perturbation defined in \cite{LiPersistentCurrent}.

\begin{figure}
\includegraphics[width=9cm]{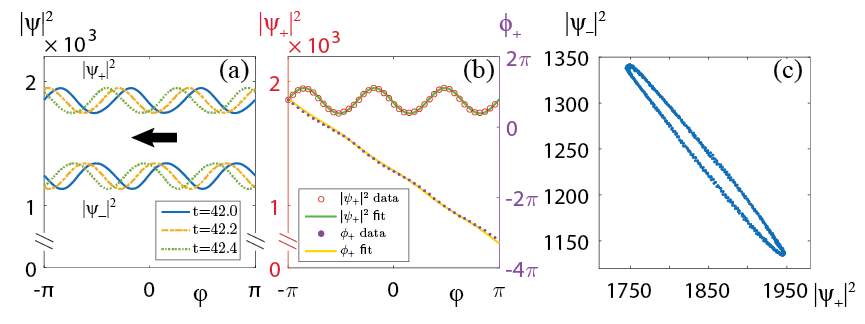} 
\caption{(Color online). (a) Azimuthal distribution of density for $\psi_{+}$ taken along the dashed-ring in Fig.~\ref{STstates}(a) at different times. (b) Fitted plots for the azimuthal density and phase distributions of $\psi_{+}$ at $t=42.0$ (see text). (c) Density dependency between two circularly polarized components at $t=42.0$. }
\label{Fit_data}
\end{figure}

To verify cnodial wave rotations, Fig.~\ref{Fit_data}(a) shows time series of the condensate density recorded along the dashed-ring, which demonstrates the periodic rotation around the center of the condensate, with the period at about $T_P \sim 10$ ps, at the frequency of terahertz. Fig.~\ref{Fit_data}(b) shows instantaneous density and phase distribution for $\psi_{+}$ along the dashed-ring, as well as plots fitted by using the expression for cnoidal waves derived in \cite{KanamotoPRA09,SmyrnakisPRA13},
\begin{equation}
|\psi_{+} (\varphi)|^2 \sim  \mathrm{cn}^2(\tilde{\varphi},k) \quad \text{and} \quad \phi_{+} (\varphi)       \sim  \mathrm{\Pi} (\xi, \tilde{\varphi},k),
\end{equation}
where $\mathrm{cn}$ and $\mathrm{\Pi}$ are Jacobi elliptic function and incomplete elliptic integral of the third kind respectively, $\tilde{\varphi}=j\,\mathrm{K}(k)\,(\varphi-\varphi_0)/\pi $ is the reduced azimuthal coordinate with $j$ the number of density dips, $\varphi_0$ a constant phase shift, and $\mathrm{K}(k)$ the complete elliptic integral of the first kind, where $k \in [0,1]$, and $\xi$ are fitting parameters. In contrast to \cite{SmyrnakisPRA13}, instead of linear dependency, the densities of the other component are related by the elliptic relation: 
$ (|\psi_{+} (\varphi)|^2)^2 /a +  (|\psi_{-} (\varphi)|^2)^2 /b=1 $, where the numerical coefficients $a$ and $b$ define the axes of the ellipse (translated to the origin) shown in Fig.~\ref{Fit_data}(c). 


\begin{figure}
\includegraphics[width=8cm]{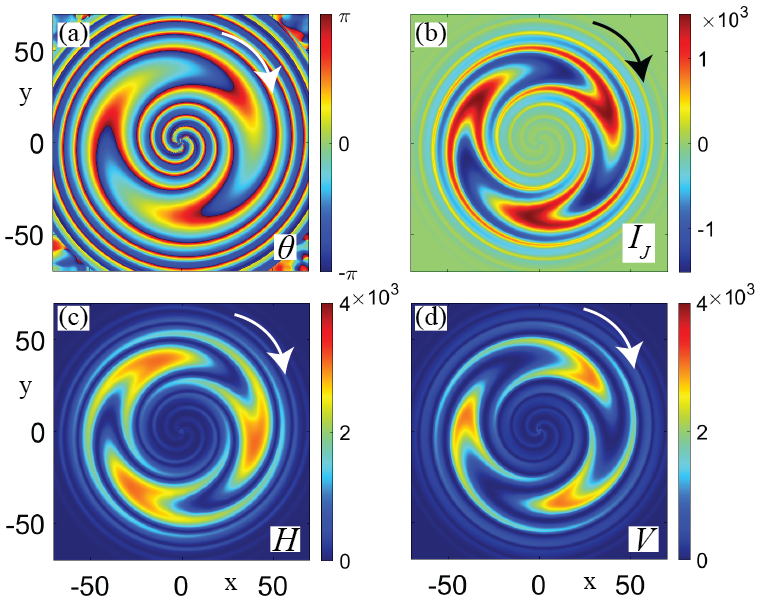} 
\caption{(Color online). Properties of the ST state shown in Fig.~\ref{STstates}. (a) and (b) Spatial distributions of the relative phase $\theta$ and the Josephson current $I_J$. (c) and (d) Density distributions in the linearly polarized basis for the horizontal $H$ and the vertical $V$ component. Arrows indicate rotation direction.}
\label{ST_Linear_Basis}
\end{figure}

The phase winding number difference between two polarization components gives rise to circulating internal Josephson currents that form the Josephson vortex \cite{KaurovPRA05,SuPRL13,SuPRA15}. The expression of the internal Josephson current for polariton systems is given by \cite{WoutersPRB08}
\begin{equation}\label{eq:I_J}
I_J (\textbf{r}, t) = |\psi_{+}| |\psi_{-}| \sin (\theta),
\end{equation}
where $\theta$ is the relative phase defined in Eq.~(\ref{eq:theta}), and the positive value of $I_J$ indicates particle flows from the $-$ component toward the $+$ component and vice versa. Fig.~\ref{ST_Linear_Basis}(a) and (b) show the corresponding $\theta$ and $I_J$ of the ST state in Fig.~\ref{STstates}. 
Both of them are time-dependent and rotating at the same speed as the cnodial wave.
In contrast to the optical vortex azimuthon, the topological charge \cite{ThoulessBook98} of $I_J$ has the same value as the number of the density dips in the azimuthal density distribution, specifically, three in our case. Here we emphasis that, unlike azimuthons supported purely by nonlinear interactions \cite{DesyatnikovPRL05,LopezOpticsExpess05,LashkinPRA09}, the ST states we are considering are supported by the Josephson vortex given by the Josephson (linear) coupling.  

The rotating Josephson vortex enables the realization of spin waves proposed recently in Ref.~\cite{GlazovPRB15} for polariton systems. The spatial propagation of spin waves manifests itself in the linearly polarized basis \cite{OktelPRL02}. As shown in Fig.~\ref{ST_Linear_Basis}(c) and (d), in the linearly polarized basis both $H$ and $V$ components of the ST state exhibit periodic density modulation with high contrast. The modulated densities, which rotate at the same speed as the density dips in Fig.~\ref{Fit_data}(a), represent the "optical ferris wheels" \cite{FrankeOpticsExpress07} in the cavity photoluminescence. They might be applicable in the design of polariton spin switch \cite{AmoNatPho10} for ultra-fast polaritonic devices.

\section{Pattern formation}\label{Sec:Striped_States}

It is well-known that both self-interference effects and nonlinear instabilities in driven-dissipative systems can lead to formation of stationary and fluctuating patterns \cite{CrossREV93,StaliunasBook03}. 
In polariton systems, pattern formations have been observed in experiments, \emph{e.g.}, the \emph{sunflower state} \cite{ChristmannPRB12} and the \emph{self-ordered state} \cite{ManniPRB13}. 
Recently, it has been proposed that modulational instability can result in the appearance of phase defects in polariton systems \cite{TimPRB15}.
As it would be shown below, similarly, instabilities introduced by the Josephson vortex can lead to pattern formation in density and phase for a polariton condensate.

\begin{figure}
\includegraphics[width=8cm]{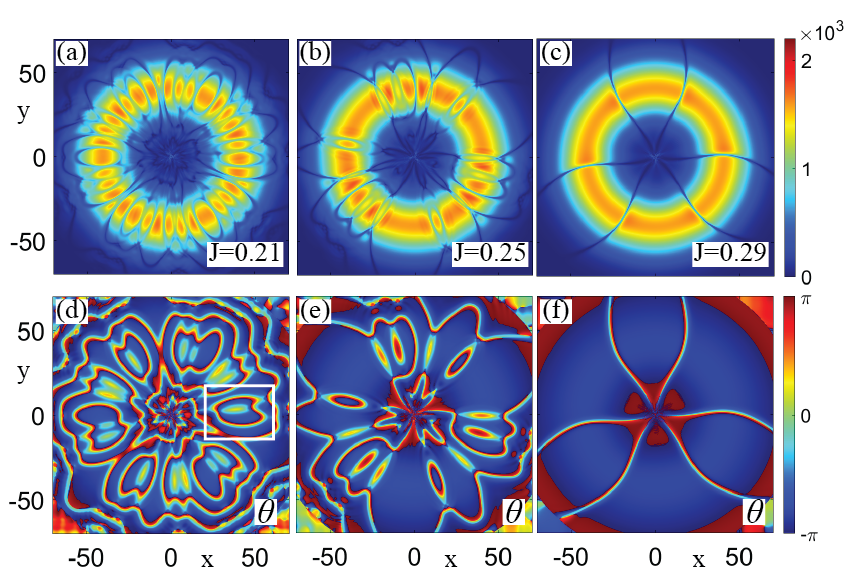} 
\caption{(Color online). (a)-(c) Density distribution of the less pumped component $|\psi_{-}|^2$ for different $J$. (d)-(f) The corresponding relative phase distribution. The white box indicates the magnification area shown in Fig.~\ref{Rotating_Stripes}(a). Other parameters are the same as in Fig.~\ref{STstates}.}
\label{Striped_States}
\end{figure}

With the increase of the linear coupling $J$, particle densities carried by the internal Josephson current become comparable to the azimuthal flow within each polarization component, so that they can break the azimuthal flow by perturbations in the form of long thin stripes with sharp phase gradients. Fig.~\ref{Striped_States} shows snap shots of non-stationary striped states. 
While both components keep their overall azimuthal flows, their density distributions are cut by density stripes that either originate from the centre of the condensate (open stripe) or form a closed loop (closed stripe).

\begin{figure}
\includegraphics[width=8cm]{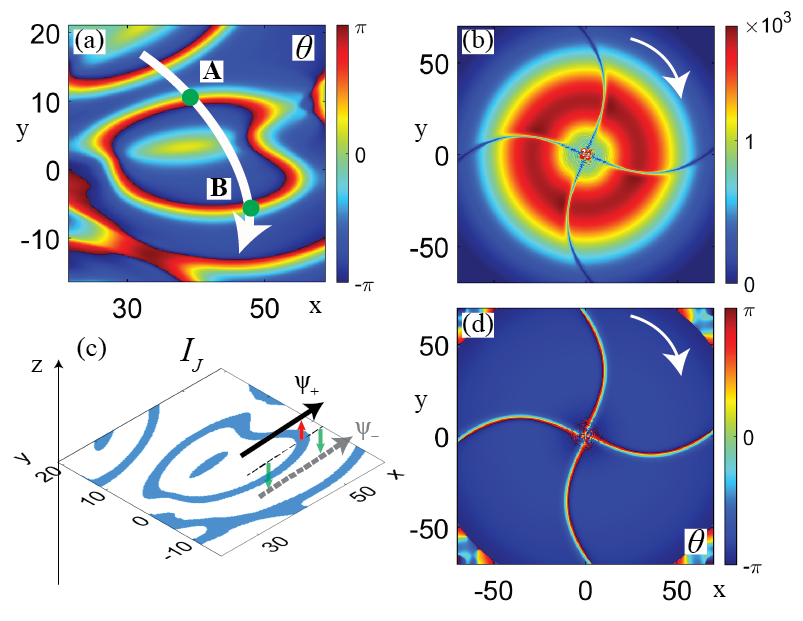} 
\caption{(Color online). (a) Magnified relative phase plot. White arrow: schematic illustration of a path having phase changes when crossing the boundary (green dots) of a closed stripe. (c) A schematic diagram showing the flow direction of the internal Josephson current (see text).
(b) and (d)  density ($|\psi_{-}|^2$) and relative phase distribution of a steadily rotating striped state. White arrows indicate the rotation direction. Parameters are the same as in Fig.~\ref{STstates} except that $m_{+}=-2$, $m_{-}=2$, and $J=0.26$.}
\label{Rotating_Stripes}
\end{figure}

The locations of the those density stripes are determined by that of the Josephson vortices, which is linked to the relative phase $\theta$.
Fig.~\ref{Striped_States}(d)-(f) show the corresponding spatial distribution of $\theta (\textbf{r})$. As we can see, the stripes are distributed randomly on top of a uniform phase background, which corresponds to a fixed relative phase $\theta_s$,  \emph{i.e.}, that of the SAB state. 
In the transverse direction of every stripe, the relative phase has a $2\pi$ phase change which consists of two $\pi$ phase changes in both polarization components separately. 
Within a given polarization component, if its particle flow is represented as a path with directions, then, when the path crosses the boundary of a stripe, there will be a $\pi$ phase change for the flow. 
Specifically, Fig.~\ref{Rotating_Stripes}(a) shows a magnified plot for the relative phase in the area highlighted by a white box in Fig.~\ref{Striped_States} and 
a schematic plot of a path crossing the closed stripe. 
There are $\pi$ phase changes with opposite sign for both components when the path passes the marked points $A$ and $B$. The overall effect is that the path has zero phase gain after crossing the boundary of a closed stripe twice. 
Therefore, the phase winding number for each polarization component will not change by passing through a closed stripe. 
In contrast, open stripes that originate from the middle of the condensate toward the outside of the pumping region such as Fig.~\ref{Striped_States}(f), can change the phase winding number.

The Josephson vortex embodied in a stripe stems from differences in particle flows within two polarization components.
Fig.~\ref{Rotating_Stripes}(c) shows a pseudo-$3D$ plot for the Josephson current $I_J$ corresponding to Fig.~\ref{Rotating_Stripes}(a). The x-y plane represents the $2D$ spatial distribution of $I_J$, where the darker and lighter color indicate the positive and negative value of $I_J$ respectively. 
Above the plane the $+$ component is placed , while below the plane there is the $-$ component.
The dark solid arrow (above the plane) indicates the particle flow in $\psi_+$ when passing through the boundary of the stripe (straight dashed line); the gray dashed arrow (below the plane) indicates the particle flow in $\psi_-$ at the same position, where both components share approximately the same flow direction as indicated. 
The difference between the speed of two flows, which in turn leads to a $2\pi$ phase change in $\theta$, results in the appearance of internal Josephson currents around the boundary of the stripe.
Whereas particles tunnel from $\psi_-$ to $\psi_+$ in the darker blue area (red arrow), the direction of the tunneling flow reverses in the lighter color area (green arrows), thus forming the Josephson vortex. Unlike the ST state and many superconducting systems, where the Josephson vortex is given by tunneling between two counter-propagating flows \cite{RoditchevNatPhys15}, 
here it is given by tunneling between two co-propagating flows. And both its amplitude and position are time-dependent.
With the continuously change of the spatial distribution of particle flows in $\psi_\pm$, the position of the closed stripe will change as a result. For the current case, Fig.~\ref{Rotating_Stripes}(a), the closed stripe will expand or merge with another closes stripe until it disappears.

Clearly, in this system the Josephson vortex has unconventional flow properties. However, we retain the terminology because of the same physical origin, \emph{i.e.}, circular exchange of particle due to the linear coupling.
 
With the further increase of the linear coupling $J$, the effect of the SAB state is more obvious. As can be seen in Fig.~\ref{Striped_States}, less density dips and relative phase stripes form with larger $J$, and the phase becomes more uniform across the pumped area. In Fig.~\ref{Striped_States}(e), the number of closed stripes reduces significantly, while keeping the number of the open stripes the same (equal to the number of density dips in Fig.~\ref{STstates}). It means that the overall phase winding number for each component can still be different from each other. In Fig.~\ref{Striped_States}(f), no closed stripes exist, and the number of the open stripes is not equal to three. In this case, both components share the same phase winding number, and such a state is very close to the SAB state. The remaining open stripes are not static. They appear and collide with each other and disappeared periodically. 

A different set of parameters supports the existence of steady rotating stripes. Figures~\ref{Rotating_Stripes}(b) and (d) show such an example. Four stripes rotated anti-clockwise steadily, with the period at about $t=96$ ps. Also, the arrangements for the $2\pi$ phase change direction in $\theta$ embodied in the stripes differs from that of Fig.~\ref{Striped_States}(f), preventing them from merging.
As the linear coupling $J$ increases further, the condensate falls into the SAB state independently of the initial winding number in each component, and the relative phase is fixed at $\theta_s$ everywhere.

Distinction should be made between defects in the relative phase of the striped states and the shock line defects in frozen states which are solutions of the complex Ginzburg-Landau equation \cite{AransonREV02,BritoPRL03}. The shock line defects are caused by the phase difference between two nearby vortices \cite{AransonREV02,BritoPRL03}, which are basically single-component phenomena for an open-dissipative system. In fact, shock lines and frozen states have been observed numerically for a single-component polariton condensate under an homogeneous incoherent pump in \cite{TimPRB15}. And it has been observed in experiment that a spiraling state called \emph{the sunflower state} \cite{ChristmannPRB12} exist under a Gaussian-shape incoherent pump, which might correspond to the frozen states after spiraling waves are established \cite{TimPRB15}. In contrast, the striped states considered here are intrinsically two-component phenomena which highly depend on the strength of the linear coupling $J$ [see Fig.~\ref{Striped_States}]. 

The competition between the nonlinear interaction and the linear coupling is ubiquitous in many multi-component nonlinear systems. Thus, the pattern formation leading to striped states could be applicable to other dissipative nonlinear systems such as atomic BECs and nonlinear optics.

\section{Spontaneous vortex generation}\label{Sec:White_noise}

In the above two sections, the orbital angular momentum of each component was imprinted by an external coherent LG pulse in the initial stage of the condensate evolution towards a steady state. While ensuring the controlled generation of angular momentum, this coherent imprinting is not essential for obtaining non-zero OAM for a polariton condensate. In fact, each component can acquire angular momentum independently starting from white noise in the process of mode selection governed by the specific spatial configuration of the incoherent pump \cite{yulin}. In this section, we will show the process of ST state generation from white noise. 
And the finial state will exhibit asymmetric density distribution due to the random capture of the initially formed vortices. 


We start discussing generally the condensate growing process that applies to a wide range of pumping configurations. 
This process, which is governed by the model equations (\ref{eq:rescaled}), is essentially a single-component phenomenon and can be illustrated by the example of a polariton condensate supported by a fully circularly polarized pump ($PL=0$). In this case the pumped polarization component, $\psi=\psi_{+}$, dominates the whole dynamical process. Before the pump reaches the threshold power, the particle density $|\psi|^2$ is zero or takes a negligibly small value. When the pumping threshold power is reached, the $|\psi|^2=0$ state is no longer dynamically stable and the condensate density will grow exponentially \cite{TauberPRX14,MatuszewskiPRB14}. If we assume that the correlation length of $\psi$ extends to the whole pumped region, then $\psi$ will grow like $\psi \sim |\psi_0| e^{i \phi_0} e^{i \omega_1\, t} e^{\omega_2 \, t}$, where $\phi_0$ is the initial phase and $\omega_{1,2}$ are the real and imaginary part of the eigenfrequency of the unstable mode, respectively \cite{LiPersistentCurrent}. This is a typical homogeneous growth scenario where the whole condensate shares the same growth rate (given by $\omega_2$), and the finial state would inherit the angular momentum completely from $\phi_0$ (under a radially symmetric pump). If $\phi_0$ has no OAM, then there is none in the finial state.

\begin{figure}
\includegraphics[width=9cm]{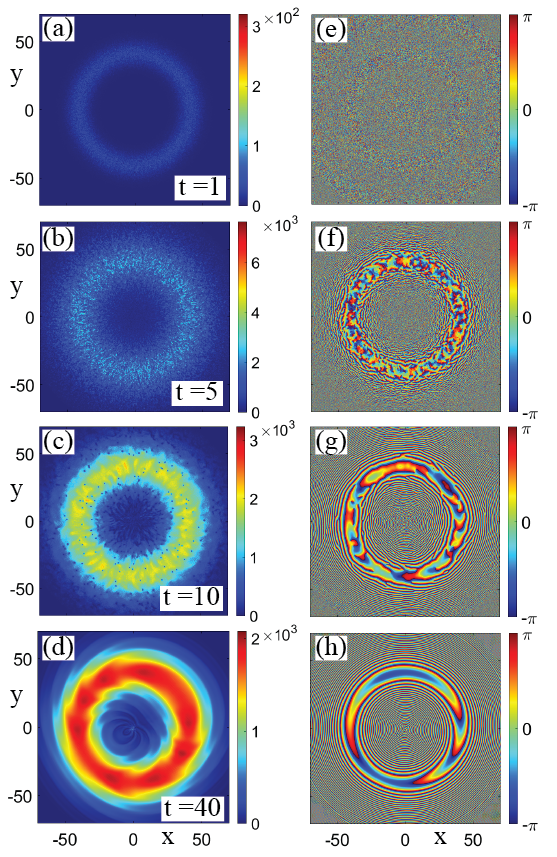} 
\caption{(Color online). Inhomogeneous growth of the condensate and spontaneous formation of a vortex state. (a)-(d) Density distribution of $|\psi_{+}|^2$. (e)-(h) Phase distribution $\phi_{+}$ for the corresponding time.}
\label{White_Noise}
\end{figure}

When noise is present in the initial state or in the driven-dissipative GP equation, the condensate will experience inhomogeneous growth and vortices will appear. A practical white noise can be generated independently at each point from a random variable $Y=n_s \, X$, where the random variable $X\sim N(0,1)$ follows the standard normal distribution \cite{noise91} and $n_s$ represents the noise strength.
Depending on the noise strength, the noise in the phase of the condensate will reduce the correlation length of the condensate, and when $n_s$ reaches a critical value, the correlation length becomes zero and thus the growth rate of the condensate will differ from point to point. Localized defects such as vortices can form during this process. 

With the continuous growth of the condensate, those initial vortices will be captured or repelled by the condensate depending on the specific pumping configuration. Fig.~\ref{White_Noise} shows snapshots of the inhomogeneous growth of the condensate and the spontaneous formation of vortices. The condensate is seeded by a sufficiently strong white noise. As we can see from Fig.~\ref{White_Noise}(a)-(c), vortices grow locally and are randomly captured within the pumped ring. Vortices with opposite charges annihilate and, if they do not cancel out completely, the remaining charge will be inherited by the bulk condensate as in Fig.~\ref{White_Noise}(d). Note that in Fig.~\ref{White_Noise} the linear Josephson coupling $J$ is set to support a ST state. The spontaneous formation of vorticity for each component can be regarded as independent from each other for the current pumping configuration. The resulted sate is an imperfect azimuthon vortex state. The same process also applies to spontaneous formation of homogeneous vortex states and striped states. 

\section{Conclusions} In conclusion, by using a dynamical mean-field model to describe two-component exciton-polariton condensate formed in the incoherent spin-polarized puming regime, we have demonstrated existence and dynamical stability of vortex azimuthons and spin patterns in an annular trapping geometry imposed by the pumping geometry. We have investigated the intrinsic connection between these nontrivial spin structures and internal Josephson currents supported by a linear polarization splitting. Our results are generally applicable to other open-dissipative systems in the context of atomic BECs and nonlinear optics.

\begin{acknowledgments}
This work was supported by the Australian Research Council (ARC) and the China Scholarship Council (CSC). The author thanks Michael Fraser, Alexey Yulin, and Elena A. Ostrovskaya for useful discussions and in-depth comments.
\end{acknowledgments}


\begin{thebibliography}{99}
\bibitem{CarusottoREV13} I. Carusotto and C. Ciuti, Rev. Mod. Phys. {\bf 85}, 299 (2013).
\bibitem{SchneiderArxiv15} C. Schneider, K. Winkler, M.D. Fraser, M. Kamp, Y. Yamamoto, E.A. Ostrovskaya, S. Hoefling, ”Exciton-Polariton Trapping and Potential Landscape Engineering”, arXiv:1510.07540 (2015).
\bibitem{ByrnesREV14} T. Byrnes, N. Y. Kim, and Y. Yamamoto, Nat. Phys. {\bf 10}, 803 (2014).

\bibitem{ManniPRL11} F. Manni, K. G. Lagoudakis, T. C. H. Liew, R. Andr\'e, and B. Deveaud-Pl\'edran, Phys. Rev. Lett. {\bf 107}, 106401 (2011).
\bibitem{DreismannPNAS14} A. Dreismann, P. Cristofolini, R. Balili, G. Christmann, F. Pinsker, N. G. Berloff, Z. Hatzopoulos, P. G. Savvidis, J. J. Baumberg, Proc. Natl. Acad. Sci. {\bf 111}, 8770 (2014).
\bibitem{LiPersistentCurrent} G. Li, M. D. Fraser, A. Yakimenko, and E. A. Ostrovskaya, Phys. Rev. B {\bf 91}, 184518 (2015).
\bibitem{SnokePNAS15} G. Liu, D. W. Snoke, A. Daley, L. N. Pfeiffer, and K. West, Proc. Natl. Acad. Sci. {\bf 112}, 2676 (2015).
\bibitem{LagoudakisScience09} K. G. Lagoudakis, T. Ostatnick\'{y}, A. V. Kavokin, Y. G. Rubo, R. Andr\'{e}, B. Deveaud-Pl\'{e}dran, Science {\bf 326}, 974 (2009).
\bibitem{KammannPRL12} E. Kammann, T. C. H. Liew, H. Ohadi, P. Cilibrizzi, P. Tsotsis, Z. Hatzopoulos, P. G. Savvidis, A. V. Kavokin, and P. G. Lagoudakis, Phys. Rev. Lett. {\bf 109}, 036404 (2012).
\bibitem{HivetNatPhys12}  R. Hivet,	H. Flayac,	D. D. Solnyshkov,	D. Tanese,	T. Boulier,	D. Andreoli,	E. Giacobino,	J. Bloch,	A. Bramati,	G. Malpuech, and A. Amo, Nat. Phys. {\bf 8}, 724 (2012).
\bibitem{ManniNatComm13} F. Manni, Y. L\'eger, Y.G. Rubo, R. Andr\'e, and  B. Deveaud, Nat. Comms. {\bf 4}, 2590 (2013).
\bibitem{Swirls15} P. Cilibrizzi, H. Sigurdsson, T. C. H. Liew, H. Ohadi, S. Wilkinson, A. Askitopoulos, I. A. Shelykh, and P. G. Lagoudakis, Phys. Rev. B {\bf 92}, 155308 (2015).
\bibitem{Shelykh05} I. A. Shelykh, A. V. Kavokin, and G. Malpuech, Phys. Stat. Solid. b {\bf 242}, 2271 (2005).
\bibitem{Kavokin05} A. Kavokin, G. Malpuech, and M. Glazov, Phys. Rev. Lett. {\bf 95}, 136601 (2005).

\bibitem{BerloffPRB10} M. O. Borgh, J. Keeling, and N. G. Berloff, Phys. Rev. B {\bf 81}, 235302 (2010).
\bibitem{LiSpin15} G. Li, T. C. H. Liew, O. A. Egorov, and E. A. Ostrovskaya, Phys. Rev. B {\bf 92}, 064304 (2015).

\bibitem{FrankeOpticsExpress07} S. Franke-Arnold, J. Leach, M. J. Padgett, V. E. Lem-bessis, D. Ellinas, A. J. Wright, J. M. Girkin, P. \"Ohberg, and A. S. Arnold, Opt. Express. {\bf 15} 8619 (2007).

\bibitem{WoutersPRL07} M. Wouters and I. Carusotto, Phys. Rev. Lett. {\bf 99}, 140402 (2007).




\bibitem{Vladimirova10} M. Vladimirova, S. Cronenberger, D. Scalbert, K. V. Kavokin, A. Miard, A. Lema\^{i}tre, J. Bloch, D. Solnyshkov, G. Malpuech, and A. V. Kavokin, Phys. Rev. B {\bf 82}, 075301 (2010).
\bibitem{GlazovPRB15} M. M. Glazov and A. V. Kavokin, Phys. Rev. B {\bf 91}, 161307(R) (2015).

\bibitem{parameters} In the numerical calculations, the following values of parameters are chosen as experimentally accessible:
$m_{LP}=10^{-4}\:m_e$ (where $m_e$ is the free electron mass), $u_a=6 \times 10^{-3}\:\text{meV}\: \mu \text{m}^2$, $u_b=-0.1u_a$, $g_R=2u_a$, 
$\gamma_c = 0.33 \:\text{ps}^{-1}$, $\gamma_R=1.5\gamma_c$, $R=0.01\: \mu \text{m}^2\: \text{ps}^{-1}$, and $J \approx 0.01\!\sim\! 0.5\: \text{meV}$.
So, the dimensionless quantities are: $u_a=7.7 \times 10^{-3}$, $u_b=-7.7 \times 10^{-4}$, $g_R=1.5\times 10^{-2}$, $\gamma_R=1.5$, $R=8.4\times 10^{-3}$, and $J\approx 0.01\!\sim\! 0.5$. This choice of parameters corresponds to an assumption of a very short lifetime of reservoir excitons, however the results do not depend on the particular value of $\gamma_R$.
\bibitem{Bao13} W. Bao and Y. Cai, Kinet. Relat. Mod., {\bf 6}, 1 (2013).

\bibitem{RuboPRL07} Yuri G. Rubo, Phys. Rev. Lett. {\bf 99}, 106401 (2007).
\bibitem{RuboPRB14} M. Toledo-Solano, M. E. Mora-Ramos, A. Figueroa, and Y. G. Rubo, Phys. Rev. B {\bf 89}, 035308 (2014).

\bibitem{ManniNatComm12} F. Manni, K. G. Lagoudakis, T. C. H Liew, R. Andr\'{e}, V. Savona, and B. Deveaud, Nat. Comms. {\bf 3}, 1309 (2012).

\bibitem{WoutersPRB08} M. Wouters, Phys. Rev. B {\bf 77}, 121302(R) (2008).

\bibitem{SanvittoNatPhys10} D. Sanvitto, F. M. Marchetti, M. H. Szyma\'{n}ska, G. Tosi, M. Baudisch, F. P. Laussy, D. N. Krizhanovskii, M. S. Skolnick, L. Marrucci, A. Lema\^{i}tre, J. Bloch, C. Tejedor, and L. Vi\~{n}a, Nat. Phys. {\bf 6}, 527 (2010).

\bibitem{DesyatnikovPRL05} A. S. Desyatnikov, A. A. Sukhorukov, and Yu. S. Kivshar, Phys. Rev. Lett. {\bf 95}, 203904 (2005).
\bibitem{LopezOpticsExpess05} S. Lopez-Aguayo, A. S. Desyatnikov, and Yu. S. Kivshar, Opt. Express {\bf 14}, 7903 (2006).
\bibitem{LashkinPRA09} V. M. Lashkin, E. A. Ostrovskaya, A. S. Desyatnikov, and Yu. S. Kivshar, Phys. Rev. A {\bf 80}, 013615 (2009).
\bibitem{MinovichOptExpress09} A. Minovich, D. N. Neshev, A. S. Desyatnikov, W. Krolikowski, and Yu. S. Kivshar, Opt. Express {\bf 17}, 23610 (2009).
\bibitem{IzdebskayaOptExpress11} Ya. V. Izdebskaya, A. S. Desyatnikov, G. Assanto, and Yu. S. Kivshar, Opt. Express {\bf 19}, 21457 (2011).

\bibitem{LobanovOptExp11} V. E. Lobanov, Y. V. Kartashov, V. A. Vysloukh, and L. Torner, Opt. Lett. {\bf 36}, 85 (2011)
\bibitem{BorovkovaOptExp11} O. V. Borovkova, V. E. Lobanov, Y. V. Kartashov, and L. Torner, Opt. Lett. {\bf 36}, 1936 (2011).
\bibitem{BorovkovaPRA12} O. V. Borovkova, V. E. Lobanov, Y. V. Kartashov, and L. Torner, Phys. Rev. A {\bf 85}, 023814 (2012).
\bibitem{OstrovskayaPRA12} E. A. Ostrovskaya, J. Abdullaev, A. S. Desyatnikov, M. D. Fraser, and Y. S. Kivshar, Phys. Rev. A {\bf 86}, 013636 (2012).

\bibitem{CarrPRA00} L. D. Carr, Charles W. Clark, and W. P. Reinhardt, Phys. Rev. A {\bf 62}, 063610 (2000).
\bibitem{CarrPRA00II} L. D. Carr, Charles W. Clark, and W. P. Reinhardt, Phys. Rev. A {\bf 62}, 063611 (2000).
\bibitem{KanamotoPRA09} R. Kanamoto, L. D. Carr, and M. Ueda, Phys. Rev. A {\bf 79}, 063616 (2009).
\bibitem{SmyrnakisPRA13} J. Smyrnakis, M. Magiropoulos, G. M. Kavoulakis, and A. D. Jackson, Phys. Rev. A {\bf 87}, 013603 (2013).
\bibitem{KamchatnovPRA14} A. M. Kamchatnov, Y. V. Kartashov, P.-\'{E}. Larr\'{e}, and N. Pavloff, Phys. Rev. A {\bf 89}, 033618 (2014).

\bibitem{PinskerPRL14} F. Pinsker and H. Flayac, Phys. Rev. Lett. {\bf 112}, 140405 (2014).

\bibitem{KaurovPRA05} V. M. Kaurov and A. B. Kuklov, Phys. Rev. A {\bf 71}, 011601(R) (2005).
\bibitem{SuPRL13} S. W. Su, S. C. Gou, A. Bradley, O. Fialko, and J. Brand, Phys. Rev. Lett. {\bf 110}, 215302 (2013).
\bibitem{SuPRA15} S. W. Su, S. C. Gou, I.Kang Liu, A. S. Bradley, O. Fialko, and J. Brand, Phys. Rev. A {\bf 91}, 023631 (2015).
\bibitem{RoditchevNatPhys15} D. Roditchev,	C. Brun,	L. Serrier-Garcia,	J. C. Cuevas,	V. H. L. Bessa,	M. V. Milo\v{s}evi\'{c},	F. Debontridder,	V. Stolyarov, and T. Cren, Nat. Phys. {\bf 11}, 332 (2015).


\bibitem{AtrePRE06} R. Atre, P. K. Panigrahi, and G. S. Agarwal, Phys. Rev. E {\bf 73}, 056611 (2006).

\bibitem{ThoulessBook98} D. J. Thouless, \emph{Topological Quantum Numbers in Nonrelativistic Physics} (World Scientific, Singapore, 1998).

\bibitem{OktelPRL02} M. \"{O}. Oktel and L. S. Levitov, Phys. Rev. Lett. {\bf 88}, 230403 (2002).


\bibitem{AmoNatPho10} A. Amo, T. C. H. Liew, C. Adrados, R. Houdr\'{e}, E. Giacobino, A. V. Kavokin, and A. Bramati, Nat. Photonics {\bf 4}, 361 (2010).

\bibitem{CrossREV93} M. C. Cross and P. C. Hohenberg, Rev. Mod. Phys. {\bf 65}, 851 (1993).

\bibitem{StaliunasBook03} K. Staliunas and V.J. S\'{a}nchez-Morcillo, \emph{Transverse Patterns in Nonlinear Optical Resonators} (Springer, Berlin, 2003).

\bibitem{ChristmannPRB12} G. Christmann, G. Tosi, N. G. Berloff, P. Tsotsis, P. S. Eldridge, Z. Hatzopoulos, P. G. Savvidis, and J. J. Baumberg, Phys. Rev. B {\bf 85}, 235303 (2012).
\bibitem{ManniPRB13} F. Manni, T. C. H. Liew, K. G. Lagoudakis, C. Ouellet-Plamondon, R. Andr\'{e}, V. Savona, and B. Deveaud, Phys. Rev. B {\bf 88}, 201303(R) (2013).

\bibitem{TimPRB15} T. C. H. Liew, O. A. Egorov, M. Matuszewski, O. Kyriienko, X. Ma, and E. A. Ostrovskaya, Phys. Rev. B {\bf 91}, 085413 (2015).




\bibitem{AransonREV02} I. S. Aranson and L. Kramer, Rev. Mod. Phys. {\bf 74}, 99 (2002).
\bibitem{BritoPRL03} C. Brito, I. S. Aranson, and H. Chat\'{e}, Phys. Rev. Lett. {\bf 90}, 068301 (2003).


\bibitem{TauberPRX14} U. C. T\"{a}uber and S. Diehl, Phys. Rev. X {\bf 4}, 021010 (2014).
\bibitem{MatuszewskiPRB14} M. Matuszewski and E. Witkowska, Phys. Rev. B {\bf 89}, 155318 (2014).

\bibitem{yulin} A. Yulin, A. S. Desyatnikov, and E. A. Ostrovskaya, 'Spontaneous mode selection in optically-induced traps for exciton-polariton condensates', in preparation.

\bibitem{noise91} Athanasios Papoulis, \emph{Probability, Random Variables and Stochastic Processes, 3rd} (McGraw-Hill, NY, 1991).

\end{thebibliography}
\end{document}